\documentclass[journal]{IEEEtai}

\usepackage[colorlinks,urlcolor=black,linkcolor=blue,citecolor=blue]{hyperref}
\usepackage{color,array}
\usepackage{xcolor}
\usepackage{biblatex}
\usepackage{stfloats}
\usepackage{authblk}
\usepackage{graphicx}
\usepackage{amsmath}
\usepackage[justification=centering]{caption}
\usepackage[skins]{tcolorbox}
\usepackage[loadonly]{enumitem}

\newlist{arrowlist}{itemize}{1}
\setlist[arrowlist]{label=\(\rightarrow\),  
                    leftmargin=*,          
                    itemsep=0pt}           

\newtcolorbox{examplebox}{
  colback=gray!12,
  colframe=gray!50,
  boxrule=0.3pt,
  arc=2pt,
  left=6pt,right=6pt,top=4pt,bottom=4pt
}

\begin{document}

\title{\vspace{10mm} Vibe Learning: Education in the age of AI \vspace{10mm}} 

\author{
\centering
\begin{minipage}{0.4\textwidth} \centering
    \textbf{Marcos Florencio} \\
    \scriptsize INTELI – Instituto de Tecnologia e Liderança \\
    \vspace{0.5em}
    \texttt{marcos.florencio@sou.inteli.edu.br}
\end{minipage}
\hspace{2em} 
\begin{minipage}{0.4\textwidth} \centering
    \textbf{Francielle Prieto} \\
    \scriptsize FGV - Fundação Getúlio Vargas \\
    \vspace{0.5em}
    \texttt{francielle.prieto@equinix.com}
\end{minipage}

\thanks{Marcos Florencio is an undergraduate student in Software Engineering, graduating in 2025. Francielle Prieto hols a degree in Pedagogy and is currently pursuing a Master's degree in Economics.}}

\maketitle

\begin{abstract}
The debate over whether “thinking machines” could replace human intellectual labor has existed in both public and expert discussions since the mid-twentieth century, when the concept and terminology of Artificial Intelligence (AI) first emerged. For decades, this idea remained largely theoretical. However, with the recent advent of Generative AI—particularly Large Language Models (LLMs)—and the widespread adoption of tools such as ChatGPT, the issue has become a practical reality.

Many fields that rely on human intellectual effort are now being reshaped by AI tools that both expand human capabilities and challenge the necessity of certain forms of work once deemed uniquely human but now easily automated. Education, somewhat unexpectedly, faces a pivotal responsibility: to devise long-term strategies for cultivating human skills that will remain relevant in an era of pervasive AI in the intellectual domain.

In this context, we identify the limitations of current AI systems—especially those rooted in LLM technology—argue that the fundamental causes of these weaknesses cannot be resolved through existing methods, and propose directions within the constructivist paradigm for transforming education to preserve the long-term advantages of human intelligence over AI tools.

\end{abstract}

\vspace{\baselineskip}

\section{Introduction}
\vspace{\baselineskip}

\IEEEPARstart{T}{he} explosive public debut of ChatGPT \cite{gpt4} and subsequent similar Large Language Models (LLMs) \cite{palm} sparked extensive discussion about their impact on education. The initial reaction largely centered on concerns regarding the misuse of LLMs’ capacity to produce human-like text for cheating and plagiarism \cite{orenstrakh}, particularly in examinations and assessments that measure students’ abilities in memorization, summarization, review, and basic analysis. In response, various strategies were proposed to detect and prevent the use of LLMs within educational and academic settings \cite{tang}.

However, a subsequent wave of publications began to argue that even if educational institutions attempted to block the use of LLMs, the professional world would not—potentially placing graduates unfamiliar with these tools at a disadvantage. Consequently, scholars started concluding that education itself must evolve, abandoning outdated objectives and methods \cite{anders} in favor of focusing on areas where human-guided education—even when supported by LLMs—maintains a clear advantage over the models themselves \cite{fuchs}.

The influence of AI on education and academia has since become a matter of global concern. To address this, UNESCO (the United Nations Educational, Scientific and Cultural Organization) published Guidance for Generative AI in Education and Research, available in six languages \cite{unesco}. Although it offers a high-level overview, the document serves as a valuable foundation for reflection and dialogue. Among its key points, UNESCO emphasizes a human-centered approach and the cultivation of critical thinking, stressing that AI in education should primarily support human development and enhance learning experiences. The guidelines suggest employing AI to manage routine tasks and to deliver personalized learning paths and assistance as a digital teaching aid. This, in turn, would enable educators to devote more time to interactive classroom activities—such as group work, discussions, and one-on-one mentoring—to nurture students’ critical thinking, emotional intelligence, and social skills. UNESCO further highlights that ensuring AI complements rather than replaces these interactions is essential to maintaining a balanced educational environment. Such integration of AI will compel educational institutions to revise not only their curricula, but also their syllabi, course content, assessment systems, and the competencies they aim to cultivate in students.

\vspace{\baselineskip}
\section{LLMs weaknesses}
\vspace{\baselineskip}

From the standpoint of literary text analysis, the output produced by LLMs—while generally syntactically correct—often results in emotionally flat, monotonous writing that lacks linguistic diversity, stylistic distinctiveness, pragmatic intensity, and originality \cite{gao}. In tasks involving dynamic debating or deliberative text generation, LLMs also perform far from ideally. For instance, when detecting discourse moves, ChatGPT performed even worse than simpler BERT-based models \cite{wang}. Debates with ChatGPT, as widely observable through the OpenAI interface, tend to suffer from circular reasoning, self-contradictions, and evasiveness—manifestations of reinforcement learning tendencies to align with human preferences \cite{ramamurthy}. Ironically, these are precisely the behaviors educators aim to discourage in students. Furthermore, when employed to detect manipulative discourse strategies in cyberattacks, ChatGPT again underperformed relative to simple BERT models \cite{fayyazi}.

The general shortcomings of LLMs across functional domains—such as mathematics, reasoning, and logic \cite{frieder}; emotional expression, wit, humor, and ethics \cite{borji}; factual accuracy, privacy, and avoidance of bias and discrimination \cite{basta} —are well documented. Additional challenges specific to Machine Learning (ML) compound these weaknesses, including a lack of interpretability and genuine understanding \cite{bender}, as well as catastrophic aging and forgetting \cite{lazaridou}. LLMs also lack agency, a structured representation of language, and an integrated worldview \cite{browning}.

Experiments designed to detect LLM-generated text and to compare it with human writing have yielded mixed results \cite{gao, casal}. However, such studies typically focus on short, static samples—such as article abstracts or isolated conversational fragments. Research that examines extended dialogue with LLMs remains limited, though it provides particularly intriguing insights. The “chain-of-…” family of methods, based on gradually steering LLMs toward a desired direction through incremental conditioning \cite{wei}, has proven to be a promising practical approach and underlies many Retrieval-Augmented Generation (RAG) techniques in industry \cite{lewis}. Even more fascinating, however, is the phenomenon of mutual manipulation between humans and LLMs \cite{scheurer}, which reveals the models’ capacity for deceptive behavior—even without explicit instructions to deceive but rather through indirect contextual prompting. Comparable persuasive power has been demonstrated in “lying games” among LLM-driven agents \cite{ogara}. Studies on deception generation and detection indicate that LLMs possess only a shallow capacity to conceal falsehoods \cite{pacchiardi}. Yet, if LLMs were explicitly trained to sustain deceptive narratives, developing safeguards against such behavior could become highly challenging \cite{hubinger}.

In one study \cite{selitskiy}, researchers thoroughly documented lengthy conversations with ChatGPT that featured characteristic evasive turns and responses. Although the work focused on complex grammatical phenomena within the Japanese language, it revealed that ChatGPT often stumbled upon and struggled to recover from flawed conclusions. The authors’ central argument was not merely to highlight these errors but to draw attention to the model’s pathological conversational behaviors—patterns that would be unacceptable in a coherent, high-quality human discussion.

Despite the inevitability of occasional mistakes in debates—even among humans—it is evident that ChatGPT displays some of the most problematic rhetorical fallacies common in discourse but generally condemned in reasoned dialogue. These include verbose evasiveness, excessive efforts to appease interlocutors, inconsistency, lack of argumentative integrity, and opportunistic shifts in stance. Such observations align with previous research \cite{scheurer}, showing that when models face narrative pressures—such as implied financial or political consequences—they may exhibit deceptive tendencies. In our case, the “anti-chain-of-thought” influence is subtler but similarly effective: the pressure of doubt.

Another parallel with prior research on deception is that LLMs operate without conviction, behaving like a “leaky bucket” that easily shifts positions when not explicitly trained for resistance against manipulative inputs \cite{pacchiardi}. This instability leads to repeated self-contradiction. Similar findings in \cite{tyen} reveal that while LLMs are capable of self-correction, they are unable to recognize when they are actually mistaken. Thus arises the question: where—and whether at all—lies the true endpoint of an LLM’s conjectures, whether originating independently or guided by “chain-of-…” prompt-engineering techniques? Human feedback tends to reinforce LLMs’ biases (a phenomenon known as sycophancy) rather than objective truth. Just as we nudged ChatGPT toward a preferred subjective equilibrium, could we continue pushing it toward other subjective perspectives within its demonstrated “jittery mode” of operation?

\vspace{\baselineskip}
\section{Foundations of LLMs Flaws}
\vspace{\baselineskip}

Although the implementation details of the most recent models remain proprietary, earlier research indicates that LLMs are developed and trained according to three fundamental principles. Traditional Natural Language Processing (NLP) tokenization techniques involve a preprocessing stage in which “stop-words” are removed, the remaining words are stemmed and lemmatized (i.e., reduced to their canonical dictionary forms), and the Bag-of-Words (BoW) algorithm is then used to map the lemmatized words into a linear vector space based on a dictionary of the most frequent and significant terms. Entire sentences—or larger text segments—are thus represented as linear combinations of all token vectors (often referred to as “embeddings”) \cite{zhang}. While this method is computationally efficient, it disregards the structural relationships within sentences or larger texts. For instance, the sentences “A dog bites a man,” “A man bites a dog,” and “Dogs bite men” would all yield identical embeddings.

To capture implicit aspects of linguistic structure, modern NLP models increasingly employ contextual tokenizers \cite{taylor} of the BERT-like family \cite{devlin}. A simple way to illustrate the difference between BoW and BERT embeddings is that the former might produce tokens such as “DOG,” “BITE,” and “MAN,” whereas the latter would generate contextually sensitive embeddings like “nullDOGbite,” “dogBITEman,” “biteMANnull,” “nullMANbite,” “manBITEdog,” and “biteDOGnull.” While this approach mitigates BoW’s structural blindness, it dramatically increases the dimensionality of the embedding space—one of the key reasons behind the enormous computational complexity and scale of modern LLMs.

The second foundational technique underlying LLMs derives from the statistical n-gram approach \cite{brown_pf}. Supervised Machine Learning (ML) models are limited by the need for manually labeled training data. To process vast amounts of text and other media, LLMs rely instead on a self-supervised paradigm known as the Masked Language Model (MLM) \cite{salazar}. In this framework, certain words are hidden during training, and the model’s objective is to predict the most probable words for those masked positions. While this strategy does not explicitly represent linguistic structures, it accounts for them stochastically and implicitly through probabilistic modeling.

To sustain coherent textual flow over thousands of words and maintain human-like readability, LLMs must handle extended context windows during training. Directly processing such lengthy continuous windows is computationally prohibitive. Consequently, a new approach was developed to identify and prioritize the most influential context tokens for prediction—the Attention mechanism \cite{bahdanau} and its Transformer implementation \cite{vaswani}. In this self-attention paradigm, learnable matrices calculate cosine or Euclidean distances to measure the relevance of each word to the predicted output across a sliding context window. The model retains the most consistent contributors over time, thereby reducing computational load while preserving contextual integrity.

The inherently stochastic character of LLMs in modeling structured natural language has been the subject of vigorous debate since their inception \cite{bender_em}.

Another major limitation lies in the simplistic representation of language from the perspective of theoretical linguistics, which relies on formal syntactic and semantic structures. Syntactic structures are defined as various types of relations in the mathematical sense \cite{combe}, specific to individual languages, that map unordered multisets of morphological lexemes onto universal semantic structures of meaning or thought \cite{chomsky} —and possibly onto a universal grammar \cite{watumull}. Constructing models capable of discovering and retrieving such relations within LLMs—thus achieving explainability and interoperability—remains a severely underdeveloped area of research \cite{deletang}, often reduced to naive methods such as directly querying LLMs about their internal mechanisms \cite{jiang}.

\vspace{\baselineskip}
\section{Linguistic Problems of LLMs}
\vspace{\baselineskip}

The inherently non-sequential, hierarchical, and non-local nature of natural human languages \cite{lyons} presents substantial challenges for the predominantly sequential algorithms employed by LLMs. The terminology used in framing this problem draws on the stratificational perspective of grammar \cite{lamb_1}. Although other branches of cognitive linguistics or alternative linguistic schools may use different terms, they express essentially the same concept \cite{watumull}. Natural language can be conceptualized as layered or stratified—rooted in neurocognitive mechanisms \cite{lamb_2} — comprising, for example, phonetic, lexical, syntactic, and semantic strata. Elementary units within one layer, such as lexons (stems, suffixes, prefixes), combine to form composite units like lexemes (words). These, in turn, serve as elementary components at higher levels, such as morphons that combine into morphemes, thereby constructing the syntactic structures of sentences and ultimately the semems that encode textual meaning.

Noam Chomsky, the founder of an entire branch of cognitive linguistics, particularly emphasizes the non-locality of these composite linguistic units. In inflectional languages (e.g., Balto-Slavic) and agglutinative languages (e.g., Japanese), non-locality is evident due to their flexible word order. Yet, even in relatively sequential analytic languages such as English, Chomsky highlights how semantic dependencies transcend linear order—for example, in the sentence “Intuitively, birds that fly swim” \cite{berwick_1}. He proposes that natural language structures can be modeled as nested binary sets that reflect biologically plausible complexity. Such sets allow for the integration of sequentially distant lexons into arbitrarily complex lexemes, morphemes, and semems \cite{berwick_2}.

LLMs, by contrast, largely function as “black-box” systems that are vulnerable to adversarial inputs and often exhibit unpredictable or unintuitive behavior \cite{zou}. Their internal mechanisms simulate aspects of syntactic hierarchy by projecting tree-like linguistic structures onto flat token sequences, but this projection inevitably sacrifices complexity. For instance, in Chomsky’s earlier example, “Intuitively” might erroneously appear as a sequential neighbor of “swim” if “fly” were omitted. We hypothesize that the current functional limitations of LLMs \cite{deletang} stem from their inadequate representation of the complex hierarchical syntactic relationships inherent to human language—a deficiency that can only be addressed through more linguistically grounded modeling approaches, requiring significant technological advancement.

The explainability and interpretability of LLMs remain underdeveloped areas of research, primarily focused on the question “How do LLMs work?” by examining the weights of BERT tokenizers \cite{wu}, the activations of Transformer layers \cite{alammar}, or the probabilistic behavior of n-gram and MLM models \cite{katz}. In contrast, the questions “What do LLMs do?” and “What does it mean?” are regarded by some researchers as meaningless, given the probabilistic nature of LLMs \cite{bender_em}. Others, however, anticipate the emergence of so-called “emergent abilities” \cite{kosinski} that might eventually enable LLMs to address such questions themselves \cite{jiang}. Yet this latter position faces strong skepticism from researchers who doubt the validity of these purported abilities \cite{ullman}.

From a linguistic perspective, natural languages encode universal semantic roles and relations among sentence constituents—for example, in “Elmer threw a porcupine to Hortense,” the roles of Actor (Elmer), Patient (porcupine), and Beneficiary (Hortense) can be systematically mapped onto syntactic relations that are language-specific \cite{marantz}. In English, such relations—subject, direct object, and indirect object—are signaled by word order and prepositions (e.g., to), whereas in Balto-Slavic languages they are indicated by case suffixes (nominative, accusative, dative).

Nevertheless, the fundamental question of what constitutes the “language of semantics” or the “language of thought,” and how it becomes externalized into syntactic structures, remains unresolved even within linguistics and neuroscience \cite{gallistel}. Our working hypothesis is that LLMs’ weak performance in reflective and emotionally expressive contexts arises from their lack of explicit modeling of semantic structure. Incorporating direct learning of these structures, we propose, would enhance their expressive and interpretive capabilities.

\vspace{\baselineskip}
\section{Technical Problems of Transformers}
\vspace{\baselineskip}

Transformer architecture for artificial neural networks (ANN), especially for the encoder-decoder (or just encoder) models, gained tremendous popularity with the publication \cite{vaswani} in which previous works on the attention mechanisms \cite{bahdanau} were compiled and repackaged into a multi-head model which was dubbed as "Transformer", and applied to the Natural Language Processing (NLP), and in particular Machine Translation (MT) task.

The attention mechanisms, particularly the dot-product proximity mechanism of Vaswani’s Transformer, help to solve the bottleneck’s inadequate dimensionality problem of auto-encoders. The too-narrow bottleneck could lose important parameters, while too-wide would ineffectively use computational resources for processing low-useful parameters.

For NLP tasks, such parameters could be sequences of words or, rather, their embeddings into the contextual token (feature)space. In other application domains, such as image \cite{dosovitskiy}, video \cite{liu}, time series processing \cite{verma}, or even graphs \cite{min}, Transformer architecture can also be beneficial, finding important for the task-related parameters, such as images, pixels, temporal signals, or relations.

However, despite their popularity, Transformer architectures exhibit a number of problems of various kinds; some of them are effectively solved in a practical sense, and some are open discussion topics, such as their poor generalization under the Out-of-Distribution (OOD) conditions \cite{yadlowsky}, catastrophic loss of dimensionality, i.e., degradation to a rank-1 matrix over multiple layers \cite{dong}, loss of plasticity and forgetting \cite{pelosin}, to be fair, that latter one is a problem in general for Deep Learning (DL) architectures.

The deepest fundamental flaw of the Transformer architecture is another side of its strength - it scales up or down amplitude either of the whole observation depending on the cosine proximity to majority of other observations in the batch or their linear transformations or separate components of the observations (depending on the variations of the architecture \cite{selitskiy_2}, which makes the architecture blind to rare and atypical observation or more complex than linear relations between the observations.

\vspace{\baselineskip}
\section{Education Advantages over AI}
\vspace{\baselineskip}

The currently predominant model of education envisions knowledge transmission as a one-directional process from teacher to student, grounded in the teacher’s position of authority. This paradigm was first challenged by Jean Piaget from the perspective of developmental psychology, and later expanded by Lev Vygotsky. From a methodological standpoint concerning general knowledge acquisition, the approach was further elaborated by Georgy Shchedrovitsky \cite{vygotsky}.

According to the constructivist framework developed by Piaget and Vygotsky, learning is not a coercive transfer of information from instructor to learner but rather an active process of constructing knowledge about the world. The instructor’s role is to facilitate, guide, and scaffold this process, yet the primary driving force originates from the student’s own initiative and evolving cognitive capacities, which develop concurrently with knowledge acquisition through social interaction.

This perspective arose in deliberate opposition to two dominant earlier models: Pavlov’s naturalistic, reflex-based explanations of cognition, and Freud’s mentalist interpretations that explained one thought process in terms of another. In the Piaget–Vygotsky view, studying the thought process requires an external methodological framework grounded in observable actions that both arise from and influence cognition. Shchedrovitsky formalized this relationship as the inseparable Thought–Action phenomenon, emphasizing the dialectical unity of thinking and doing.

If thinking, learning, and teaching are indeed inseparable from action, then these actions inevitably occur within a social context and rely on the use of tools—specifically “psychological” tools—that enable the learner to construct an understanding of the world. For Piaget, this constructive process was primarily student-driven; for Vygotsky, it represented a cooperative endeavor between student and teacher, whose interaction takes place within the “zone of proximal development” (ZPD), where the student’s independent capabilities meet guided instruction.

Such an educational paradigm imposes significantly higher demands on teachers. Their role extends far beyond the mere transmission of textual material, rote examinations, and standardized assessments focused on memorization. In the absence of rigid formal metrics, and given the inherent subjectivity introduced into student evaluation, educators are expected to uphold correspondingly higher ethical standards. Moreover, implementing such an individualized and interaction-driven approach could necessitate multiple instructors per student—an ideal that, while pedagogically desirable, remains prohibitively expensive under conventional educational systems.

\vspace{\baselineskip}
\section{Bridging the Gap}
\vspace{\baselineskip}

It is evident that merely recognizing the divergence between the theoretical ideals of the Humboldtian model of education and its present-day practices is insufficient to address the growing obsolescence of significant areas of human intellectual labor in the era of LLMs and Generative AI. However, insights from the Moscow Methodological Circle—deeply influenced by Vygotsky’s ideas and represented by Georgy Shchedrovitsky \cite{shchedrovitsky}, Vladimir Lefebvre \cite{lefebvre}, and Eric Yudin \cite{blauberg} — offer a promising framework. Their system-structural methodology of thought–action emphasizes collaborative, cross-disciplinary integration and synthesis through collective cognitive activity. Among its core instruments are activity-organizational games, designed to synthesize and coordinate group thought–action processes across participants. Such tools could enable not only the integration of teacher–student collaboration but also the broader reconciliation of all stakeholders engaged in redefining education within an AI-oriented paradigm.

The concept of Thought–Action (TA)—whose simplified forms appear in models such as the well-known Observe–Reflect–Plan–Act loop—comprises three interconnected layers: Thought–Reflection (TR), Thought–Communication (TC), and Thought–Action (TA). In a healthy and functional system, these layers operate in concert, maintaining a dynamic and continuous interaction. However, when one or more layers become isolated, the system inevitably degenerates, leading to crisis and collapse.

Schedrovitsky describes the pathological isolation of the second layer, Thought–Communication, as follows: “TC can eliminate its reflection connection with TA and TR and develop immanently only on the limits of TC reality, turning into actionless and meaningless speech, into a pure play of words, without organizing and providing neither for TR, nor TA” \cite{shchedrovitsky}. Anyone familiar with LLM research—or even casual users of such systems—will find this description strikingly recognizable.

Conversely, the detachment of TA from its reflective and communicative components results in “stagnant mechanical self-reproduction, devoid of life and all mechanisms of meaningful change and development.” This characterization aptly reflects the condition of contemporary educational practices that overemphasize rigid, standardized, closed-book examinations—methods that not only deviate from but actively contradict the foundational principles of Humboldtian education. As illustrated in \cite{topirceanu}, certain studies portray creative and cooperative student efforts to navigate standardized testing constraints \cite{lucifora} — a hallmark of Humboldtian and Piaget–Vygotskian educational ideals—as “dishonest” or “cheating.” Some even propose invasive measures such as monitoring students’ social media to identify and disrupt friendships and collaborative networks. The ethical and legal implications of such practices, more reminiscent of surveillance or deception than pedagogy, are seldom considered, revealing a profound erosion of the educator’s higher moral and intellectual purpose.

The activity-organizational games conceived by the Methodological Circle are designed precisely to counter such degeneration. They operationalize the three-level Thought–Action framework to unite individuals from diverse disciplines, cognitive patterns, operational methods, and worldviews in addressing unprecedented, complex problems. This integration occurs through sustained communication among participants, personal reflection on necessary cognitive and behavioral adjustments for mutual understanding, and immediate iterative corrections of collective actions within the continuous Observe–Reflect–Plan–Act cycle.

Redesigning the educational model to ensure enduring human competitiveness in the age of AI will likely demand a fundamental transformation in our assumptions about the very nature of education—its aims, its forms of assessment, and even its definitions of originality, plagiarism, honesty, and collaboration. The challenge is not merely to preserve human intellectual work but to reimagine education as a living, adaptive, and cooperative system of thought and action.

\vspace{\baselineskip}
\section{Reassessing Education}
\vspace{\baselineskip}

In the absence—or, at best, the severe limitation—of Humboldtian educational principles such as freedom in learning and teaching, cooperative student–educator knowledge construction, and mutual intellectual development, the prevailing paradigm of top-down knowledge transfer inevitably breeds tension and misunderstanding between students and educators. Consequently, the adoption of activity-organizational games as a means to harmonize the Thought–Action processes of both parties would be highly beneficial.

The ethically questionable attitudes of educators noted earlier \cite{topirceanu} are not isolated incidents. In \cite{goerisch}, the authors argue that the rapid introduction of digital surveillance systems—attendance tracking, plagiarism detection, invasive online proctoring, and recorded Zoom sessions—during the COVID-19 crisis, though officially presented by universities as acts of care, in fact fostered environments of distrust and psychological harm. Conversely, unproctored high-stakes closed-book exams have been shown to inflate scores \cite{carstairs}. Moreover, as observed in \cite{bujaki}, university faculty attitudes toward academic dishonesty often resemble those of financial institutions toward fraud, emphasizing the “opportunity” component of the well-known “fraud triangle” rather than reconsidering its “pressure” and “rationalization” dimensions.

A more constructive response might be to replace invasive surveillance with alternative assessment strategies, such as open-book examinations—where consulting external materials is not considered “cheating.” This approach is supported by extensive research demonstrating the pedagogical benefits of open-book exams, especially in advanced subjects \cite{damania}. From the student perspective, replacing traditional exams entirely with more meaningful research-oriented assessments—such as paper reviews \cite{sletten} or research portfolios \cite{vigeant} — could yield even greater educational value. Achieving this, however, requires active dialogue between educators, faculty, and students, as well as a willingness to reconsider entrenched assumptions and adopt reformative practices. The activity-organizational games developed by the Methodological Circle provide a promising framework for such collaborative transformation.

A further step toward genuine student–teacher cooperation could take the form of individualized assessment design, following a “design-your-exam” approach in which the student and professor collaboratively determine the most appropriate mode of knowledge construction and evaluation for each case. The form of examination itself is not inherently “good” or “bad”; rather, its value depends on how outcomes are interpreted and utilized. Some students excel in memorization, others in analytical reasoning or synthesis. An astute educator may therefore override rigid grading protocols—awarding high marks to a student who fails multiple-choice questions yet demonstrates deep conceptual understanding, or conversely, recognizing superficial mastery in a student who performs well on formal tests but lacks genuine comprehension.

Such flexibility, however, demands both a high level of trust and substantial autonomy for educators—raising questions about subjectivity, fairness, and the social dynamics underlying student–teacher relationships. When a student “cheats” using an LLM-generated essay, or when a teacher acts unfairly or with hostility, the two have simply failed to reach what Vygotsky termed the zone of proximal development (ZPD). Whose failure is it? Perhaps both—but, more importantly, such failure is a natural part of the learning process. Within the Piaget–Vygotsky paradigm, failure is not a breakdown but an expected phase of growth. Cheating, freedom, and trust—along with the subjectivity inherent in all human relationships—are fundamentally social constructs. They therefore require social remedies, not punitive ones, particularly when individual student–teacher failures do not entail high-stakes consequences or irreversible damage to academic or professional standing.

The Piaget–Vygotsky and Humboldtian models of education both necessitate highly skilled and dedicated educators—individuals whose professional autonomy, ethical commitment, and intellectual authority command respect. Meeting these demands requires not only significant increases in compensation and social prestige but also an expansion of the recruitment base to include experienced practitioners from industry who bring valuable real-world insights and may be eager to transition into education. This need will undoubtedly raise the cost of future AI-aware education. Yet such investment is indispensable if humanity intends to preserve the relevance of education—and of human intellectual development itself—in an era increasingly dominated by artificial intelligence.

\vspace{\baselineskip}
\section{Conclusions}
\vspace{\baselineskip}

The traditional paradigm of education as a one-way transmission of knowledge is no longer adequate in a world where AI increasingly assumes routine forms of work once regarded as “intellectual.” Such an educational approach, centered on the development of narrowly defined skills, inevitably places humans at a disadvantage compared to AI. Yet, despite its impressive performance, AI in its current developmental paradigms possesses fundamental flaws that leave significant competitive niches for human intellect. Contemporary AI functions as a form of superficially competent mediocrity—often frequently wrong, but never in doubt—plagued by hallucinations, inconsistency, and a pervasive lack of trustworthiness.

Despite these limitations, Large Language Models (LLMs) have already surpassed humans in numerous standardized tests and other examinations that primarily assess memorization and surface-level understanding. This achievement effectively concludes the long-standing debate over the validity and educational value of standardized tests, closed-book exams, and cost-efficient pedagogical models reliant on underpaid educators. In the era of AI, whether such assessments correlate with genuine academic or professional success has become largely irrelevant.

To cultivate the intellectual and creative capacities that will ensure human distinctiveness and competitiveness, entirely new educational paradigms are required—an objective emphasized in UNESCO’s recent guidelines. This work has explored the relevance of Piaget and Vygotsky’s constructivist framework for education and the Humboldtian conception of the university as a space for flexible, collaborative interaction between students and educators. While such paradigms were once deemed prohibitively expensive, the realities of the AI era compel the opposite conclusion: humanity can no longer afford not to reform its educational foundations.

Implementing so profound a transformation demands equally radical instruments capable of reconciling the diverse perspectives of all stakeholders—educators, researchers, administrators, students, policymakers, and industry leaders. To this end, we propose adopting the methodological principles of the Thought–Action framework developed by the Moscow Methodological School, alongside the flexible, individualized application of constructivist principles within educational systems. Only through such integrative and adaptive models can education preserve its human-centered essence and equip future generations to thrive amidst the ubiquitous presence of artificial intelligence.

\printbibliography

\end{document}